\documentclass[twoside,english,preprint,amsmath,nofootinbib]{revtex4}
\usepackage{mathrsfs}
\usepackage[T1]{fontenc}
\usepackage[latin9]{inputenc}
\usepackage{amsmath}
\usepackage{amssymb}
\usepackage{esint}

\makeatletter
\@ifundefined{textcolor}{}
{%
 \definecolor{BLACK}{gray}{0}
 \definecolor{WHITE}{gray}{1}
 \definecolor{RED}{rgb}{1,0,0}
 \definecolor{GREEN}{rgb}{0,1,0}
 \definecolor{BLUE}{rgb}{0,0,1}
 \definecolor{CYAN}{cmyk}{1,0,0,0}
 \definecolor{MAGENTA}{cmyk}{0,1,0,0}
 \definecolor{YELLOW}{cmyk}{0,0,1,0}
 }

\@ifundefined{definecolor}{\@ifundefined{definecolor}{\@ifundefined{definecolor}{\@ifundefined{definecolor}{\@ifundefined{definecolor}{\@ifundefined{definecolor}{\@ifundefined{definecolor}{\@ifundefined{definecolor}{\@ifundefined{definecolor}
 {\usepackage{color}}{}
}{}}{}}{}}{}}{}}{}}{}}{}\usepackage[dvipdfm,bookmarks=true]{hyperref}

\makeatother

\usepackage{babel}

\makeatother

\usepackage{babel}

\makeatother

\usepackage{babel}

\makeatother

\usepackage{babel}

\makeatother

\usepackage{babel}

\makeatother

\usepackage{babel}

\makeatother

\usepackage{babel}

\makeatother

\usepackage{babel}

\makeatother

\usepackage{babel}

\makeatother

\usepackage{babel}

\makeatother

\usepackage{babel}

\makeatother

\usepackage{babel}

\begin{document}

\title{{\LARGE Leading-Order Actions of Goldstino Fields}}

\date{\today}

\author{Haishan Liu}

\email{liu@zimp.zju.edu.cn}

\author{Hui Luo}

\email{huiluo@zimp.zju.edu.cn}

\author{Mingxing Luo}

\email{luo@zimp.zju.edu.cn}

\author{Liucheng Wang}

\email{liuchengwang@zimp.zju.edu.cn} 
\thanks{(Corresponding Author)}

\affiliation{Zhejiang Institute of Modern Physics, Department of Physics, Zhejiang
University, Hangzhou, Zhejiang 310027, P.R.China}
\begin{abstract}
This paper starts with a self-contained discussion of the so-called Akulov-Volkov action ${\cal S}_{\rm AV}$,
which is traditionally taken to be the leading-order action of Goldstino field.
Explicit expressions for ${\cal S}_{\rm AV}$ and its chiral version ${\cal S}_{\rm AV}^{\rm ch}$ are presented.
We then turn to the issue on how these actions are related to the leading-order action ${\cal S}_{\rm NL}$
proposed in the newly proposed constrained superfield formalism.
We show that ${\cal S}_{\rm NL}$ may yield ${\cal S}_{\rm AV}$/${\cal S}_{\rm AV}^{\rm ch}$ or a totally different action ${\cal S}_{\rm KS}$,
depending on how the auxiliary field in the former is integrated out.
However, ${\cal S}_{\rm KS}$ and ${\cal S}_{\rm AV}$/${\cal S}_{\rm AV}^{\rm ch}$ always yield the same $S$-matrix elements,
as one would have expected from general considerations in quantum field theory.
\end{abstract}
\maketitle

Supersymmetry (SUSY) is arguably among the most attractive extensions
of the standard model. It renders a reasonable framework to circumvent
the hierarchy problem and has interesting phenomenological implications
at the TeV scale. Tremendous efforts have been made on the subject
in the last several decades. Hopefully, it is to be discovered in
the coming LHC experiments.

To be consistent with existing experiments and to have certain predictive
power, SUSY must be broken and broken spontaneously. According to
the general theory of spontaneously global symmetry breaking, this
would result in a massless neutral Nambu-Goldstone fermion, the Goldstino.%
\footnote{ In supergravity, the Goldstino is absorbed by the gravitino particle
and becomes the $\pm1/2$ helicity components of the latter. However,
if the SUSY breaking scale is much smaller than the Planck scale,
the lower energy physics of gravitino will be dominated by the Goldstino.
In a sense, this provides a supersymmetric version of the equivalence
theorem. Therefore, it makes sense to investigate the physics of Goldstino
independently, as it may provide an interesting window to look into
SUSY.%
}

For its low energy physics, the Goldstino can be studied in the
framework of nonlinear realization of SUSY. The leading-order action
of Goldstino field was traditionally taken to be the so-called
Akulov-Volkov action ${\cal S}_{{\rm AV}}$ \cite{AV} or its chiral
version ${\cal S}_{{\rm AV}}^{{\rm ch}}$ \cite{SamuelWess}. Both
actions are manifestly invariant under nonlinear SUSY
transformations. In the newly proposed constrained superfield
formalism, the leading-order action of Goldstino field is assumed to
be one ${\cal S}_{{\rm NL}}$  \cite{Seiberg}. In this paper, we will
show that ${\cal S}_{{\rm NL}}$ may yield ${\cal S}_{{\rm
AV}}$/${\cal S}_{{\rm AV}}^{{\rm ch}}$ or a totally different action
${\cal S}_{{\rm KS}}$, depending on how the auxiliary field in the
former is integrated out. ${\cal S}_{{\rm KS}}$ takes a particularly
simple form, but doe not have transparent properties under nonlinear
SUSY transformations. However, ${\cal S}_{{\rm KS}}$, ${\cal
S}_{{\rm AV}}$ and ${\cal S}_{{\rm AV}}^{{\rm ch}}$ always yield the
same $S$-matrix elements, regardless how the auxiliary field is
integrated out, as one would have expected from general
considerations in quantum field theory.

In the standard (non-chiral) version of nonlinear realization of SUSY,
the Goldstino field $\lambda$ is assumed to change nonlinearly under
SUSY transformations \cite{AV,WessBagger}
\begin{equation}
\left\{ \begin{array}{l}
\delta_{\xi}\lambda_{\alpha}=\frac{1}{\kappa}\xi_{\alpha}-i\kappa(\lambda\sigma^{\mu}\bar{\xi}-\xi\sigma^{\mu}\bar{\lambda})\partial_{\mu}\lambda_{\alpha},\\
\delta_{\xi}\bar{\lambda}_{\dot{\alpha}}=\frac{1}{\kappa}\bar{\xi}_{\dot{\alpha}}-i\kappa(\lambda\sigma^{\mu}\bar{\xi}-\xi\sigma^{\mu}\bar{\lambda})\partial_{\mu}\bar{\lambda}_{\dot{\alpha}},
\end{array}\right.\label{eq:nonchiral goldstino}
\end{equation}
 while matter fields $\zeta$ are to change according to \cite{IK1,IK2}
\begin{equation}
\delta_{\xi}\zeta=-i\kappa(\lambda\sigma^{\mu}\bar{\xi}-\xi\sigma^{\mu}\bar{\lambda})\partial_{\mu}\zeta.
\end{equation}
 The Akulov-Volkov action assumes the following form \cite{AV,WessBagger}
\begin{equation}
{\cal S}_{{\rm AV}}=-\frac{1}{2\kappa^{2}}\int d^{4}x\;\det\emph{T},\label{eq:AV1}
\end{equation}
 where $\emph{T}\,_{\mu}^{\nu}=\delta_{\mu}^{\nu}-i\kappa^{2}\partial_{\mu}\lambda\sigma^{\nu}\bar{\lambda}+i\kappa^{2}\lambda\sigma^{\nu}\partial_{\mu}\bar{\lambda}$.
It is invariant under the SUSY transformation Eq (\ref{eq:nonchiral goldstino})
since the change of $\det\emph{T}$ is a total derivative
\[
\delta_{\xi}\det\emph{T}=-i\kappa\partial_{\mu}[(\lambda\sigma^{\mu}\bar{\xi}-\xi\sigma^{\mu}\bar{\lambda})\det\emph{T}\;].
\]
 Expanding $\det\emph{T}$ in terms of $\kappa$ explicitly,
\begin{eqnarray}
\det\emph{T} & = & 1-i\kappa^{2}(\partial_{\mu}\lambda\sigma^{\mu}\bar{\lambda}-\lambda\sigma^{\mu}\partial_{\mu}\bar{\lambda})\label{eq:det T}\\
 &  & -\kappa^{4}\left[i\epsilon^{\mu\nu\rho\gamma}\lambda\sigma_{\rho}\bar{\lambda}\partial_{\mu}\lambda\sigma_{\gamma}\partial_{\nu}\bar{\lambda}+\bar{\lambda}^{2}\partial_{\mu}\lambda\sigma^{\mu\nu}\partial_{\nu}\lambda+\lambda^{2}\partial_{\mu}\bar{\lambda}\bar{\sigma}^{\mu\nu}\partial_{\nu}\bar{\lambda}\right]\nonumber \\
 &  & -i\kappa^{6}\lambda^{2}\bar{\lambda}\left[\bar{\sigma}^{\rho}\partial_{\rho}\lambda\partial_{\mu}\bar{\lambda}\bar{\sigma}^{\mu\nu}\partial_{\nu}\bar{\lambda}+2\bar{\sigma}^{\nu}\partial_{\mu}\lambda\partial_{\nu}\bar{\lambda}\bar{\sigma}^{\rho\mu}\partial_{\rho}\bar{\lambda}\right]\nonumber \\
 &  & -i\kappa^{6}\bar{\lambda}^{2}\lambda\left[\sigma^{\rho}\partial_{\rho}\bar{\lambda}\partial_{\mu}\lambda\sigma^{\mu\nu}\partial_{\nu}\lambda+2\sigma^{\nu}\partial_{\mu}\bar{\lambda}\partial_{\nu}\lambda\sigma^{\rho\mu}\partial_{\rho}\lambda\right].\nonumber
\end{eqnarray}

Noticing that the $\kappa^{8}$ terms are absent in the above expression,
in contrast with \cite{AV,LLW2}. This was first observed in \cite{Kuzenko}
and reconfirmed recently in \cite{Zheltukhin}. Here we provide another
verification by a brute force calculation. According to \cite{AV,LLW2},
the $\kappa^{8}$ terms are proportional to
\[
\partial_{\mu}\bar{\lambda}\bar{\sigma}^{\mu\nu}\partial_{\nu}\bar{\lambda}\partial_{\rho}\lambda\sigma^{\rho\gamma}\partial_{\gamma}\lambda+\partial_{\mu}\bar{\lambda}\bar{\sigma}^{\nu\gamma}\partial_{\rho}\bar{\lambda}\partial_{\nu}\lambda\sigma^{\mu\rho}\partial_{\gamma}\lambda+4\partial_{\mu}\bar{\lambda}\bar{\sigma}^{\mu\rho}\partial_{\nu}\bar{\lambda}\partial_{\rho}\lambda\sigma^{\gamma\nu}\partial_{\gamma}\lambda.
\]
 Since these terms come from the determinant of a $4\times4$ matrix,
possible nonvanishing terms are only those with spacetime derivatives
of different Lorentz indices. We may take $\partial_{1}\bar{\lambda}$,
$\partial_{2}\bar{\lambda}$, $\partial_{3}\lambda$ and $\partial_{0}\lambda$
for example. All relevant terms are in the following
\begin{eqnarray*}
 &  & 4\partial_{1}\bar{\lambda}\bar{\sigma}^{12}\partial_{2}\bar{\lambda}\partial_{3}\lambda\sigma^{30}\partial_{0}\lambda+4\partial_{1}\bar{\lambda}\bar{\sigma}^{30}\partial_{2}\bar{\lambda}\partial_{3}\lambda\sigma^{12}\partial_{0}\lambda+4\partial_{1}\bar{\lambda}\bar{\sigma}^{13}\partial_{2}\bar{\lambda}\partial_{3}\lambda\sigma^{02}\partial_{0}\lambda\\
 &  & +4\partial_{1}\bar{\lambda}\bar{\sigma}^{10}\partial_{2}\bar{\lambda}\partial_{0}\lambda\sigma^{32}\partial_{3}\lambda+4\partial_{2}\bar{\lambda}\bar{\sigma}^{23}\partial_{1}\bar{\lambda}\partial_{3}\lambda\sigma^{01}\partial_{0}\lambda+4\partial_{2}\bar{\lambda}\bar{\sigma}^{20}\partial_{1}\bar{\lambda}\partial_{0}\lambda\sigma^{31}\partial_{3}\lambda,
\end{eqnarray*}
 which can be regrouped as
\[
\underset{j=1,2,3}{\sum}\left(i\partial_{1}\bar{\lambda}\sigma^{j}\partial_{2}\bar{\lambda}\partial_{3}\lambda\sigma^{j}\partial_{0}\lambda-i\partial_{1}\bar{\lambda}\sigma^{j}\partial_{2}\bar{\lambda}\partial_{3}\lambda\sigma^{j}\partial_{0}\lambda\right).
\]
It vanishes trivially. All other terms can be worked out similarly.

The action ${\cal S}_{{\rm AV}}$ in Eq (\ref{eq:AV1}) can also be
constructed with the help of superfield formalism by promoting the
Goldstino field $\lambda$ to a superfield $\Lambda$ \cite{WessBagger}
\begin{equation}
\Lambda=\exp(\theta Q+\bar{\theta}\bar{Q})\times\lambda.
\end{equation}
 An invariant action can be obtained by taking the $D$-component
of $\bar{\Lambda}^{2}\Lambda^{2}$ \cite{WessBagger}, namely
\begin{equation}
{\cal S}_{{\rm AV}}=-\frac{\kappa^{2}}{2}\int d^{4}xd^{4}\theta\;\bar{\Lambda}^{2}\Lambda^{2}.\label{eq:AV2}
\end{equation}
 Expanding $\Lambda$ in terms of $\theta$ and $\bar{\theta}$, one
reproduces Eq (\ref{eq:AV1}). On the other hand, one notices that the superfield Goldstino
$\kappa\Lambda(x)=\theta'=\theta+\kappa\lambda(z)$, where $z=x-i\kappa\lambda(z)\sigma\bar{\theta}+i\kappa\theta\sigma\bar{\lambda}(z)$.
This procedure of changing variables from $(x,\theta,\bar{\theta})$ to $(z,\theta',\bar{\theta}')$ was pioneered in \cite{IK1,IK2}.
Changing the integration variables, one has \cite{LLW}
\begin{equation}
{\cal S}_{{\rm AV}}=-\frac{\kappa^{2}}{2}\int d^{4}zd^{4}\theta'\det\emph{T}\det\emph{M}\left(\frac{\bar{\theta}'}{\kappa}\right)^{2}\left(\frac{\theta'}{\kappa}\right)^{2}=-\frac{1}{2\kappa^{2}}\int d^{4}z\det\emph{T},
\end{equation}
where $\det\emph{T}\det\emph{M}$ is the Jacobian determinant of this
transformation and $\emph{M}\,_{\mu}^{\nu}=\delta_{\mu}^{\nu}-i\kappa\theta\sigma^{\nu}\bar{\lambda}_{\mu}+i\kappa\lambda_{\mu}\sigma^{\nu}\bar{\theta}$.
Explicitly,
\begin{eqnarray}
\det\emph{M} & = & 1+i\kappa(\lambda_{\mu}\sigma^{\mu}\bar{\theta}-\theta\sigma^{\mu}\bar{\lambda}_{\mu})\label{eq:det M}\\
 &  & -\kappa^{2}\left[i\epsilon^{\mu\nu\rho\gamma}\theta\sigma_{\rho}\bar{\theta}\lambda_{\mu}\sigma_{\gamma}\bar{\lambda}_{\nu}+\bar{\theta}^{2}\lambda_{\mu}\sigma^{\mu\nu}\lambda_{\nu}+\theta^{2}\bar{\lambda}_{\mu}\bar{\sigma}^{\mu\nu}\bar{\lambda}_{\nu}\right]\nonumber \\
 &  & +i\kappa^{3}\theta^{2}\bar{\theta}\left[\bar{\sigma}^{\rho}\lambda_{\rho}\bar{\lambda}_{\mu}\bar{\sigma}^{\mu\nu}\bar{\lambda}_{\nu}+2\bar{\sigma}^{\nu}\lambda_{\mu}\bar{\lambda}_{\nu}\bar{\sigma}^{\rho\mu}\bar{\lambda}_{\rho}\right]\nonumber \\
 &  & +i\kappa^{3}\bar{\theta}^{2}\theta\left[\sigma^{\rho}\bar{\lambda}_{\rho}\lambda_{\mu}\sigma^{\mu\nu}\lambda_{\nu}+2\sigma^{\nu}\bar{\lambda}_{\mu}\lambda_{\nu}\sigma^{\rho\mu}\lambda_{\rho}\right],\nonumber
\end{eqnarray}
 where $\lambda_{\mu}=(\emph{T}^{-1})_{\mu}^{\nu}\partial_{\nu}\lambda$.
In \cite{LLW2}, there were $\kappa^{4}\theta^{2}\bar{\theta}^{2}$
terms proportional to
\[
\bar{\lambda}_{\mu}\bar{\sigma}^{\mu\nu}\bar{\lambda}_{\nu}\lambda_{\rho}\sigma^{\rho\gamma}\lambda_{\gamma}+\bar{\lambda}_{\mu}\bar{\sigma}^{\nu\gamma}\bar{\lambda}_{\rho}\lambda_{\nu}\sigma^{\mu\rho}\lambda_{\gamma}+4\bar{\lambda}_{\mu}\bar{\sigma}^{\mu\rho}\bar{\lambda}_{\nu}\lambda_{\rho}\sigma^{\gamma\nu}\lambda_{\gamma}.
\]
 They can be shown to vanish by the same line of arguments for the
$\kappa^{8}$ terms in $\det\emph{T}$.

For discussions related to chiral superfields, it is convenient to
introduce an alternative (chiral) Goldstino field $\tilde{\lambda}$
\cite{Zumino}. Under SUSY transformations, $\tilde{\lambda}$
is to change as
\begin{equation}
\left\{ \begin{array}{l}
\delta_{\xi}\tilde{\lambda}_{\alpha}=\frac{1}{\kappa}\xi_{\alpha}-2i\kappa\tilde{\lambda}\sigma^{\mu}\bar{\xi}\partial_{\mu}\tilde{\lambda}_{\alpha},\\
\delta_{\xi}\bar{\tilde{\lambda}}_{\dot{\alpha}}=\frac{1}{\kappa}\bar{\xi}_{\dot{\alpha}}+2i\kappa\xi\sigma^{\mu}\bar{\tilde{\lambda}}\partial_{\mu}\bar{\tilde{\lambda}}_{\dot{\alpha}}.
\end{array}\right.\label{eq:chiral goldstino}
\end{equation}
 $\tilde{\lambda}$ is not a new nonlinear realization of SUSY. It
is related to $\lambda$ via \cite{SamuelWess,IK1}
\begin{equation}
\begin{array}{l}
\ \ \ \ \ \ \tilde{\lambda}_{\alpha}(x)=\lambda_{\alpha}(z),\ \ \ \ \ \ \ \ \ \ \ \ \ \ \ \bar{\tilde{\lambda}}_{\dot{\alpha}}(x)=\bar{\lambda}_{\dot{\alpha}}(z^{*}),\\
z=x-i\kappa^{2}\lambda(z)\sigma\bar{\lambda}(z),\ \ \ \ \ \ z^{*}=x+i\kappa^{2}\lambda(z^{*})\sigma\bar{\lambda}(z^{*}).
\end{array}\label{eq:2 goldstinos}
\end{equation}
 Explicit relations between $\lambda$ and $\tilde{\lambda}$ can
be obtained by iterations as
\begin{eqnarray}
\lambda_{\alpha} & = & \tilde{\lambda}_{\alpha}+i\kappa^{2}\tilde{\upsilon}^{\mu}\partial_{\mu}\tilde{\lambda}_{\alpha}-\kappa^{4}\tilde{\upsilon}^{\mu}\partial_{\mu}\tilde{\lambda}\sigma^{\nu}\bar{\tilde{\lambda}}\partial_{\nu}\tilde{\lambda}_{\alpha}+\kappa^{4}\upsilon^{\mu}\tilde{\lambda}\sigma^{\nu}\partial_{\mu}\bar{\tilde{\lambda}}\partial_{\nu}\tilde{\lambda}_{\alpha}\label{eq:2 lambda}\\
 &  & -\frac{1}{2}\kappa^{4}\tilde{\upsilon}^{\mu}\tilde{\upsilon}^{\nu}\partial_{\mu}\partial_{\nu}\tilde{\lambda}_{\alpha}+\frac{i}{2}\kappa^{6}\tilde{\upsilon}^{\mu}\tilde{\upsilon}^{\nu}\partial_{\mu}\partial_{\nu}\tilde{\upsilon}^{\rho}\partial_{\rho}\tilde{\lambda}_{\alpha}+i\kappa^{6}\tilde{\upsilon}^{\mu}\partial_{\mu}\tilde{\upsilon}^{\nu}\partial_{\nu}\tilde{\upsilon}^{\rho}\partial_{\rho}\tilde{\lambda}_{\alpha},\nonumber \\
\tilde{\lambda}_{\alpha} & = & \lambda_{\alpha}-i\kappa^{2}\upsilon^{\mu}\partial_{\mu}\lambda_{\alpha}-\frac{1}{2}\kappa^{4}\upsilon^{\mu}\upsilon^{\nu}\partial_{\mu}\partial_{\nu}\lambda_{\alpha}-\kappa^{4}\upsilon^{\mu}\partial_{\mu}\upsilon^{\nu}\partial_{\nu}\lambda_{\alpha}\label{eq:2 lambda2}\\
 &  & +i\kappa^{6}\upsilon^{\mu}\partial_{\mu}\upsilon^{\nu}\partial_{\nu}\upsilon^{\rho}\partial_{\rho}\lambda_{\alpha}+\frac{i}{2}\kappa^{6}\upsilon^{\mu}\upsilon^{\nu}\partial_{\mu}\partial_{\nu}\upsilon^{\rho}\partial_{\rho}\lambda_{\alpha},\nonumber
\end{eqnarray}
 where $\upsilon\,^{\mu}=\lambda\sigma^{\mu}\bar{\lambda}$ and $\tilde{\upsilon}\,^{\mu}=\tilde{\lambda}\sigma^{\mu}\bar{\tilde{\lambda}}$.
Eq (\ref{eq:2 lambda2}) agrees with the expression in \cite{SamuelWess}
but differs from the one in \cite{SamuelWess2} by a factor of 2 in
the last term. Similar to $\Lambda$, a superfield $\tilde{\Lambda}$
could be constructed from $\tilde{\lambda}$ via \cite{SamuelWess}
\begin{equation}
\tilde{\Lambda}=\exp(\theta Q+\bar{\theta}\bar{Q})\times\tilde{\lambda},
\end{equation}
 out of which one can construct an invariant action of $\tilde{\lambda}$ \cite{SamuelWess}
\begin{equation}
{\cal S}_{{\rm AV}}^{{\rm ch}}=-\frac{\kappa^{2}}{2}\int d^{4}xd^{4}\theta\;\bar{\tilde{\Lambda}}^{2}\tilde{\Lambda}^{2}.\label{eq:SW}
\end{equation}
 Expanding ${\cal S}_{{\rm AV}}^{{\rm ch}}$ in terms of $\kappa$,
\begin{eqnarray}
{\cal S}_{{\rm AV}}^{{\rm ch}} & = & -\frac{1}{2\kappa^{2}}\int d^{4}x[1-i\kappa^{2}(\partial_{\mu}\tilde{\lambda}\sigma^{\mu}\bar{\tilde{\lambda}}-\tilde{\lambda}\sigma^{\mu}\partial_{\mu}\bar{\tilde{\lambda}})\label{eq:SW2}\\
 &  & +\kappa^{4}(2\tilde{\lambda}^{2}\partial_{\mu}\bar{\tilde{\lambda}}\bar{\sigma}^{\nu\mu}\partial_{\nu}\bar{\tilde{\lambda}}+2\bar{\tilde{\lambda}}^{2}\partial_{\mu}\tilde{\lambda}\sigma^{\nu\mu}\partial_{\nu}\tilde{\lambda}-\frac{1}{4}\tilde{\lambda}^{2}\partial^{2}\bar{\tilde{\lambda}}^{2}-\frac{1}{4}\bar{\tilde{\lambda}}^{2}\partial^{2}\tilde{\lambda}^{2}\nonumber \\
 &  & -2\tilde{\lambda}\partial_{\mu}\tilde{\lambda}\bar{\tilde{\lambda}}\partial^{\mu}\bar{\tilde{\lambda}}-2\partial_{\mu}\tilde{\lambda}\sigma^{\nu}\bar{\sigma}^{\mu}\tilde{\lambda}\bar{\tilde{\lambda}}\partial_{\nu}\bar{\tilde{\lambda}}-2\partial_{\mu}\bar{\tilde{\lambda}}\bar{\sigma}^{\nu}\sigma^{\mu}\bar{\tilde{\lambda}}\tilde{\lambda}\partial_{\nu}\tilde{\lambda}+4\tilde{\lambda}\sigma^{\mu}\partial_{\nu}\bar{\tilde{\lambda}}\partial_{\mu}\tilde{\lambda}\sigma^{\nu}\bar{\tilde{\lambda}})\nonumber \\
 &  & +i\kappa^{6}\bar{\tilde{\lambda}}^{2}(\tilde{\lambda}^{2}\partial^{2}\tilde{\lambda}\sigma^{\mu}\partial_{\mu}\bar{\tilde{\lambda}}-\partial_{\nu}\tilde{\lambda}^{2}\partial_{\rho}\bar{\tilde{\lambda}}\bar{\sigma}^{\rho}\sigma^{\nu}\bar{\sigma}^{\mu}\partial_{\mu}\tilde{\lambda}-4\tilde{\lambda}\sigma^{\rho}\bar{\sigma}^{\nu}\partial_{\rho}\tilde{\lambda}\partial_{\nu}\tilde{\lambda}\sigma^{\mu}\partial_{\mu}\bar{\tilde{\lambda}})\nonumber \\
 &  & -i\kappa^{6}\tilde{\lambda}^{2}(\bar{\tilde{\lambda}}^{2}\partial_{\mu}\tilde{\lambda}\sigma^{\mu}\partial^{2}\bar{\tilde{\lambda}}-\partial_{\nu}\bar{\tilde{\lambda}}^{2}\partial_{\mu}\bar{\tilde{\lambda}}\bar{\sigma}^{\mu}{\sigma}^{\nu}\bar{\sigma}^{\rho}\partial_{\rho}\tilde{\lambda}-4\bar{\tilde{\lambda}}\bar{\sigma}^{\rho}\sigma^{\nu}\partial_{\rho}\bar{\tilde{\lambda}}\partial_{\mu}\tilde{\lambda}\sigma^{\mu}\partial_{\nu}\bar{\tilde{\lambda}})\nonumber \\
 &  & +16\kappa^{8}\tilde{\lambda}^{2}\bar{\tilde{\lambda}}^{2}\partial_{\mu}\tilde{\lambda}\sigma^{\mu\nu}\partial_{\nu}\tilde{\lambda}\partial_{\rho}\bar{\tilde{\lambda}}\bar{\sigma}^{\rho\gamma}\partial_{\gamma}\bar{\tilde{\lambda}}].\nonumber
\end{eqnarray}
 ${\cal S}_{{\rm AV}}^{{\rm ch}}$ seems to differ from ${\cal S}_{{\rm AV}}$
drastically. In particular, there is a $\kappa^{8}$ term in ${\cal S}_{{\rm AV}}^{{\rm ch}}$.
However, one notices that $\bar{\tilde{\lambda}}^{2}\tilde{\lambda}^{2}=\bar{\lambda}^{2}\lambda^{2}$,
by a close inspection of Eq (\ref{eq:2 lambda}). One thus has
\begin{equation}
\bar{\tilde{\Lambda}}^{2}\tilde{\Lambda}^{2}=\bar{\Lambda}^{2}\Lambda^{2}.\label{eq:4 goldstino}
\end{equation}
 By taking the $\theta^{2}\bar{\theta}^{2}$ term on both side of
this equation, one readily gets ${\cal S}_{{\rm AV}}^{{\rm ch}}={\cal S}_{{\rm AV}}$.

In the newly proposed constrained superfield formalism \cite{Seiberg},
the Goldstino field is assumed to reside in the chiral superfield%
\footnote{In this paper, superfields and their components in the linear SUSY
are hatted while their counterparts in the nonlinear SUSY are not.
Other notations and conventions conform to those of \cite{WessBagger}.
All symbols can be found in \cite{LLW2}, if not explicitly defined
in this paper.%
}
\begin{equation}
\hat{X}_{{\rm NL}}=\frac{\hat{G}^{2}}{2\hat{F}}+\sqrt{2}\theta\hat{G}+\hat{F}\theta^{2},\label{eq:x}
\end{equation}
 which satisfies the constraint $\hat{X}_{{\rm NL}}^{2}=0$. As shown
in \cite{LLW2,LLZ}, this constraint on $\hat{X}_{{\rm NL}}$ and
other constraints on matter superfields can all be reformulated in
the language of the standard realization of nonlinear SUSY, provided
that one makes the following identification \cite{LLZ}
\begin{equation}
\tilde{\lambda}=\frac{\hat{G}}{\sqrt{2}\kappa\hat{F}}.\label{eq:lambdax}
\end{equation}
 Of course, $\lambda$ can then be constructed according to Eq (\ref{eq:2 lambda}).

As shown in \cite{IK1,IK2,IK3,LLW2}, spontaneously broken linear
SUSY theories can always reformulated nonlinearly if the Goldstino
field is identified \cite{LLW,LLZ}. This is based upon the following
observation: a linear superfield $\hat{\Omega}(x,\theta,\bar{\theta})$
can always be converted to a set of nonlinear matter fields, via
\begin{equation}
\Omega(x,\theta,\bar{\theta})=\exp[-\kappa\lambda(x)Q-\kappa\bar{\lambda}(x)\bar{Q}]\times\hat{\Omega}(x,\theta,\bar{\theta}),\label{eq:promotion}
\end{equation}
 where $\Omega(x,\theta,\bar{\theta})$ transforms under SUSY transformations
according to
\[
\delta_{\xi}\Omega=-i\kappa(\lambda\sigma^{\mu}\bar{\xi}-\xi\sigma^{\mu}\bar{\lambda})\partial_{\mu}\Omega.
\]
 In particular, the non-linearized $\hat{X}_{{\rm NL}}$ is $X_{{\rm NL}}=\exp\{i\theta\sigma^{\mu}\bar{\theta}\triangle_{\mu}^{+}\}F\theta^{2}$
\cite{LLZ}, where
\begin{equation}
F=-\kappa^{4}\overline{\lambda^{'}}^{2}\partial^{2}\tilde{\lambda}^{2}\hat{F}-2i\kappa^{2}\hat{F}\partial_{\mu}\tilde{\lambda}\sigma^{\mu}\overline{\lambda^{'}}+2\kappa^{4}\tilde{\lambda}\partial^{2}\tilde{\lambda}\overline{\lambda^{'}}^{2}\hat{F}\label{eq:nonlinearF}
\end{equation}
 and
\[
\overline{\lambda^{'}}=\bar{\tilde{\lambda}}-2i\kappa^{2}\tilde{\lambda}\sigma^{\mu}\bar{\tilde{\lambda}}\partial_{\mu}\bar{\tilde{\lambda}}-2\kappa^{4}\tilde{\lambda}^{2}\bar{\tilde{\lambda}}\bar{\sigma}^{\nu}\sigma^{\mu}\partial_{\nu}\bar{\tilde{\lambda}}\partial_{\mu}\bar{\tilde{\lambda}}+\kappa^{4}\tilde{\lambda}^{2}\bar{\tilde{\lambda}}^{2}\partial^{2}\bar{\tilde{\lambda}}.
\]

In \cite{Seiberg}, the leading-order action for the Goldstino field
is proposed to be%
\footnote{$\kappa^{-1}=\sqrt{2}f$, to conform with notations in \cite{Seiberg}.%
}
\begin{equation}
{\cal S}_{{\rm NL}}=\int d^{4}xd^{4}\theta\hat{X}_{{\rm NL}}^{\dagger}\hat{X}_{{\rm NL}}+\frac{1}{\sqrt{2}\kappa}\int d^{4}xd^{2}\theta\,\hat{X}_{{\rm NL}}+\frac{1}{\sqrt{2}\kappa}\int d^{4}xd^{2}\bar{\theta}\,\hat{X}_{{\rm NL}}^{\dagger}.\label{eq:KS}
\end{equation}
 Following the general procedure in \cite{LLW2,IK3}, this action
can be reexpressed as
\begin{eqnarray}
{\cal S}_{{\rm NL}} & = & \int d^{4}xd^{4}\theta\;\det\emph{T}\det\emph{M}\; e^{-i\theta\sigma^{\mu}\bar{\theta}\triangle_{\mu}^{-}}\; F^{\dagger}\bar{\theta}^{2}\; e^{i\theta\sigma^{\mu}\bar{\theta}\triangle_{\mu}^{+}}\; F\theta^{2}\\
 &  & +\frac{1}{\sqrt{2}\kappa}\int d^{4}xd^{2}\theta\;\det\emph{T}\det\emph{M}_{+}\; F\theta^{2}+\frac{1}{\sqrt{2}\kappa}\int d^{4}xd^{2}\bar{\theta}\;\det\emph{T}\det\emph{M}_{-}\; F^{\dagger}\bar{\theta}^{2},\nonumber
\end{eqnarray}
 where
\begin{equation}
\left\{ \begin{array}{l}
\det\emph{M}_{+}=1-2i\kappa\theta\sigma^{\mu}\bar{\lambda}_{\mu}+4\kappa^{2}\theta^{2}\bar{\lambda}_{\mu}\bar{\sigma}^{\nu\mu}\bar{\lambda}_{\nu},\\
\det\emph{M}_{-}=1+2i\kappa\lambda_{\mu}\sigma^{\mu}\bar{\theta}+4\kappa^{2}\bar{\theta}^{2}\lambda_{\mu}\sigma^{\nu\mu}\lambda_{\nu}.
\end{array}\right.
\end{equation}
 Integrating out the $\theta$'s, one has
\begin{equation}
{\cal S}_{{\rm NL}}=\int d^{4}x\;\det\emph{T}\;\left(F^{\dagger}F+\frac{1}{\sqrt{2}\kappa}F+\frac{1}{\sqrt{2}\kappa}F^{\dagger}\right).\label{eq:SNL0}
\end{equation}
 Being a nonlinear matter field, the auxiliary fields $F$ can be
integrated out without breaking the nonlinear SUSY, via its equation
of motion
\begin{equation}
F=-\frac{1}{\sqrt{2}\kappa}.\label{eq:auxiliary}
\end{equation}
 Substituting this $F$ back into Eq (\ref{eq:SNL0}), one recovers
the Akulov-Volkov action ${\cal S}_{{\rm AV}}$ in Eq (\ref{eq:AV1}).
On the other hand, substituting this $F$ into Eq (\ref{eq:nonlinearF}),
one gets by iterations an expression of the linear auxiliary field
$\hat{F}$ in terms of $\tilde{\lambda}$ solely%
\footnote{Similar expressions in two dimensions were presented in \cite{Zachos}.%
}
\begin{eqnarray}
\hat{F} & = & -\frac{1}{\sqrt{2}\kappa}+\sqrt{2}i\kappa\partial_{\mu}\tilde{\lambda}\sigma^{\mu}\bar{\tilde{\lambda}}+2\sqrt{2}\kappa^{3}(\partial_{\mu}\tilde{\lambda}\sigma^{\mu}\partial_{\nu}\bar{\tilde{\lambda}}\tilde{\lambda}\sigma^{\nu}\bar{\tilde{\lambda}}+\bar{\tilde{\lambda}}^{2}\partial_{\mu}\tilde{\lambda}\sigma^{\mu\nu}\partial_{\nu}\tilde{\lambda})\label{eq:F}\\
 &  & +\sqrt{2}i\kappa^{5}(\tilde{\lambda}^{2}\bar{\tilde{\lambda}}^{2}\partial_{\mu}\tilde{\lambda}\sigma^{\mu}\partial^{2}\bar{\tilde{\lambda}}-2\tilde{\lambda}^{2}\partial_{\mu}\tilde{\lambda}\sigma^{\mu}\partial_{\rho}\bar{\tilde{\lambda}}\bar{\tilde{\lambda}}\bar{\sigma}^{\nu}\sigma^{\rho}\partial_{\nu}\bar{\tilde{\lambda}}+4\bar{\tilde{\lambda}}^{2}\tilde{\lambda}\sigma^{\mu}\partial_{\mu}\bar{\tilde{\lambda}}\partial_{\rho}\tilde{\lambda}\sigma^{\rho\nu}\partial_{\nu}\tilde{\lambda})\nonumber \\
 &  & -8\sqrt{2}\kappa^{7}\tilde{\lambda}^{2}\bar{\tilde{\lambda}}^{2}\partial_{\mu}\bar{\tilde{\lambda}}\bar{\sigma}^{\mu\nu}\partial_{\nu}\bar{\tilde{\lambda}}\partial_{\rho}\tilde{\lambda}\sigma^{\rho\gamma}\partial_{\gamma}\tilde{\lambda}.\nonumber
\end{eqnarray}

Integrating out the $\theta$'s directly in Eq. (\ref{eq:KS}), one
has \cite{Seiberg}
\begin{equation}
{\cal S}_{{\rm NL}}=\int d^{4}x\left[i\partial_{\mu}\hat{G}\sigma^{\mu}\bar{\hat{G}}+\frac{\hat{\bar{G}}^{2}}{2\hat{\bar{F}}}\partial^{2}\frac{\hat{G}^{2}}{2\hat{F}}+\hat{F}^{\dagger}\hat{F}+\frac{1}{\sqrt{2}\kappa}\hat{F}+\frac{1}{\sqrt{2}\kappa}\hat{F}^{\dagger}\right].\label{eq:SNL1}
\end{equation}
 Reexpress this in terms of the nonlinear Goldstino field $\tilde{\lambda}$
via (\ref{eq:lambdax})
\begin{equation}
{\cal S}_{{\rm NL}}=\int d^{4}x\left[2i\kappa^{2}\hat{F}^{\dagger}\partial_{\mu}(\hat{F}\tilde{\lambda})\sigma^{\mu}\bar{\tilde{\lambda}}+\kappa^{4}\hat{F}^{\dagger}\bar{\tilde{\lambda}}^{2}\partial^{2}(\hat{F}\tilde{\lambda}^{2})+\hat{F}^{\dagger}\hat{F}+\frac{1}{\sqrt{2}\kappa}\hat{F}+\frac{1}{\sqrt{2}\kappa}\hat{F}^{\dagger}\right].\label{eq:SNL2}
\end{equation}
 One obtains then the equation of motion for the auxiliary field $\hat{F}$
\begin{equation}
2i\kappa^{2}\partial_{\mu}(\hat{F}\tilde{\lambda})\sigma^{\mu}\bar{\tilde{\lambda}}+\kappa^{4}\bar{\tilde{\lambda}}^{2}\partial^{2}(\hat{F}\tilde{\lambda}^{2})+\hat{F}+\frac{1}{\sqrt{2}\kappa}=0.
\end{equation}
 This results in an explicit expression for $\hat{F}$ by tedious
iterations, which is identical to the one given in Eq (\ref{eq:F}).
Substituting $\hat{F}$ back into Eq (\ref{eq:SNL2}), we find
\begin{eqnarray}
{\cal S}_{{\rm NL}} & = & -\frac{1}{2\kappa^{2}}\int d^{4}x[1-2i\kappa^{2}\partial_{\mu}\tilde{\lambda}\sigma^{\mu}\bar{\tilde{\lambda}}-4\kappa^{4}(\partial_{\mu}\tilde{\lambda}\sigma^{\mu}\partial_{\nu}\bar{\tilde{\lambda}}\tilde{\lambda}\sigma^{\nu}\bar{\tilde{\lambda}}+\bar{\tilde{\lambda}}^{2}\partial_{\mu}\tilde{\lambda}\sigma^{\mu\nu}\partial_{\nu}\tilde{\lambda})\label{eq:KS2}\\
 &  & -2i\kappa^{6}(\tilde{\lambda}^{2}\bar{\tilde{\lambda}}^{2}\partial_{\mu}\tilde{\lambda}\sigma^{\mu}\partial^{2}\bar{\tilde{\lambda}}-2\tilde{\lambda}^{2}\partial_{\mu}\tilde{\lambda}\sigma^{\mu}\partial_{\rho}\bar{\tilde{\lambda}}\bar{\tilde{\lambda}}\bar{\sigma}^{\nu}\sigma^{\rho}\partial_{\nu}\bar{\tilde{\lambda}}+4\bar{\tilde{\lambda}}^{2}\tilde{\lambda}\sigma^{\mu}\partial_{\mu}\bar{\tilde{\lambda}}\partial_{\rho}\tilde{\lambda}\sigma^{\rho\nu}\partial_{\nu}\tilde{\lambda})\nonumber \\
 &  & +16\kappa^{8}\tilde{\lambda}^{2}\bar{\tilde{\lambda}}^{2}\partial_{\mu}\bar{\tilde{\lambda}}\bar{\sigma}^{\mu\nu}\partial_{\nu}\bar{\tilde{\lambda}}\partial_{\rho}\tilde{\lambda}\sigma^{\rho\gamma}\partial_{\gamma}\tilde{\lambda}],\nonumber
\end{eqnarray}
 which is identical to ${\cal S}_{{\rm AV}}^{{\rm ch}}$ in Eq (\ref{eq:SW2})
up to total derivative terms.

Notice that $\hat{F}$ does not have definite transformation properties
in the formalism of nonlinear SUSY, an invariant action under nonlinear
SUSY transformations is not guaranteed if $\hat{F}$ is integrated
out in Eq (\ref{eq:SNL2}). The nonlinear SUSY invariance of Eq (\ref{eq:KS2})
may be largely due to the fact that Eq (\ref{eq:SNL2}) is quadratic
in $\hat{F}$. If $\hat{F}$ is integrated out via the equation of
motion obtained from Eq. (\ref{eq:SNL1}) directly, one would have
\cite{Seiberg}
\begin{equation}
\hat{F}_{{\rm KS}}=-\frac{1}{\sqrt{2}\kappa}-\frac{\kappa^{3}}{\sqrt{2}}\hat{\bar{G}}^{2}\partial^{2}\hat{G}^{2}+\frac{3\kappa^{7}}{\sqrt{2}}\hat{G}^{2}\hat{\bar{G}}^{2}\partial^{2}\hat{G}^{2}\partial^{2}\hat{\bar{G}}^{2}.
\end{equation}
 In this case, one obtains a particularly simple action from Eq (\ref{eq:SNL1})
\begin{eqnarray}
{\cal S}_{{\rm KS}} & = & \int d^{4}x\left(-\frac{1}{2\kappa^{2}}+i\partial_{\mu}\hat{\bar{G}}\bar{\sigma}^{\mu}\hat{G}+\frac{\kappa^{2}}{2}\hat{\bar{G}}^{2}\partial^{2}\hat{G}^{2}-\frac{\kappa^{6}}{2}\hat{G}^{2}\hat{\bar{G}}^{2}\partial^{2}\hat{G}^{2}\partial^{2}\hat{\bar{G}}^{2}\right).\label{eq:wrong AV}
\end{eqnarray}
 Similar to $\hat{F}$, $\hat{G}$ does not have definite transformation
properties in the formalism of nonlinear SUSY either. So, it is not
transparent how ${\cal S}_{{\rm KS}}$ changes under nonlinear SUSY
transformations. Naively, one may use Eq (\ref{eq:lambdax}) to convert
the $\hat{G}$ field to nonlinear Goldstino field $\tilde{\lambda}$.
For example, one may take $\hat{G}=\sqrt{2}\kappa\tilde{\lambda}\hat{F}_{{\rm KS}}$,
which can be easily solved by
\begin{equation}
\hat{G}=-\tilde{\lambda}-\kappa^{4}\tilde{\lambda}\bar{\tilde{\lambda}}^{2}\partial^{2}\tilde{\lambda}^{2}.
\end{equation}
 Substituting this into ${\cal S}_{{\rm KS}}$, one has
\begin{eqnarray}
 & -\frac{1}{2\kappa^{2}}\int d^{4}x(1+2i\kappa^{2}\tilde{\lambda}\sigma^{\mu}\partial_{\mu}\bar{\tilde{\lambda}}-\kappa^{4}\bar{\tilde{\lambda}}^{2}\partial^{2}\tilde{\lambda}^{2}-2i\kappa^{6}\tilde{\lambda}^{2}\partial_{\mu}\tilde{\lambda}\sigma^{\mu}\bar{\tilde{\lambda}}\partial^{2}\bar{\tilde{\lambda}}^{2}\label{eq:wrong AV2}\\
 & \hspace{6pt}+2i\kappa^{6}\bar{\tilde{\lambda}}^{2}\partial^{2}\tilde{\lambda}^{2}\tilde{\lambda}\sigma^{\mu}\partial_{\mu}\bar{\tilde{\lambda}}-3\kappa^{8}\tilde{\lambda}^{2}\bar{\tilde{\lambda}}^{2}\partial^{2}\tilde{\lambda}^{2}\partial^{2}\bar{\tilde{\lambda}}^{2}).\nonumber
\end{eqnarray}
 Unfortunately, this new form is not invariant under nonlinear SUSY
transformations. On the other hand, one may take $\hat{G}=\sqrt{2}\kappa\tilde{\lambda}\hat{F}$
with $\hat{F}$ in (\ref{eq:F}). But this does not yield an invariant
action either. This makes the point. Integrating out the $\hat{F}$
directly from ${\cal S}_{{\rm NL}}$ does not necessarily generate
an invariant action under nonlinear SUSY transformations. Consequently,
${\cal S}_{{\rm KS}}$ cannot be identified with ${\cal S}_{{\rm AV}}$
in a straightforward manner.

However, ${\cal S}_{{\rm AV}}$, ${\cal S}_{{\rm AV}}^{{\rm ch}}$,
and ${\cal S}_{{\rm KS}}$ are linked intrinsically via ${\cal S}_{{\rm NL}}$
and Eq (\ref{eq:lambdax}). They should generate the same $S$-matrix
elements, since $S$-matrix does not change under field redefinitions
and how auxiliary fields are integrated \cite{Coleman}. This can
be easily verified at the tree level, though complications arise at
loop levels due to change of measures in path integrals \cite{Weinberg}.
Given its simple structure, it could be advantageous to use ${\cal S}_{{\rm KS}}$
in practical calculations.

For illustrations, we list below the $S$-matrix elements of several
elementary processes, which are obtained from any of these actions.
For processes involving four Goldstinos, the $S$-matrix elements
can be read off from the effective operator
\begin{equation}
\mathscr{Q}_{4}=i\kappa^{2}\int d^{4}x:\bar{\psi}_{{\rm in}}^{2}(x)\partial_{\mu}\psi_{{\rm in}}(x)\partial^{\mu}\psi_{{\rm in}}(x):.
\end{equation}
 Here the $:\ :$ denotes normal ordering of operators and $\psi_{{\rm in}}$
stands for the in-state operators of $\lambda$, $\tilde{\lambda}$
and $\hat{G}$ when the actions ${\cal S}_{{\rm AV}}$, ${\cal S}_{{\rm AV}}^{{\rm ch}}$
and ${\cal S}_{{\rm KS}}$ are used respectively. Specifically, $\psi_{{\rm in}}$
is the solution of the massless Dirac equation
\[
i\bar{\sigma}^{\mu}\partial_{\mu}\psi_{{\rm in}}=0,
\]
 from which one can also obtain the free propagator
\begin{equation}
D_{\alpha\dot{\beta}}^{F}(x-y)=<0|T\psi_{\alpha}^{{\rm in}}(x)\bar{\psi}_{\dot{\beta}}^{{\rm in}}(y)|0>=\int\frac{d^{4}p}{(2\pi)^{4}}\frac{ip\cdot\sigma_{\alpha\dot{\beta}}}{p^{2}}e^{ip\cdot(x-y)},
\end{equation}
\begin{equation}
\bar{D}^{F\dot{\alpha}\beta}(x-y)=<0|T\bar{\psi}_{in}^{\dot{\alpha}}(x)\psi_{in}^{\beta}(y)|0>=\int\frac{d^{4}p}{(2\pi)^{4}}\frac{ip\cdot\bar{\sigma}^{\dot{\alpha}\beta}}{p^{2}}e^{ip\cdot(x-y)}.
\end{equation}
 For processes involving six Goldstinos, the $S$-matrix elements
can be read off from
\begin{equation}
\mathscr{Q}_{6}=-4\kappa^{4}\int d^{4}xd^{4}y:\bar{\psi}_{in}^{2}(x)\frac{\partial\psi_{{\rm in}}(x)}{\partial x^{\mu}}\frac{\partial D^{F}(x-y)}{\partial x_{\mu}}\bar{\psi}_{in}(y)\frac{\partial\psi_{{\rm in}}(y)}{\partial y^{\nu}}\frac{\partial\psi_{{\rm in}}(y)}{\partial y_{\nu}}:
\end{equation}
 while for processes involving eight Goldstinos
\begin{eqnarray}
\mathscr{Q}_{8} & = & -4i\kappa^{6}\int d^{4}xd^{4}yd^{4}z\label{eq:last}\\
 &  & :[4\bar{\psi}_{in}^{2}(x)\frac{\partial\psi_{{\rm in}}(x)}{\partial x^{\mu}}\frac{\partial D^{F}(x-y)}{\partial x_{\mu}}\bar{\psi}_{in}(y)\frac{\partial\psi_{{\rm in}}(y)}{\partial y^{\nu}}\frac{\partial D^{F}(y-z)}{\partial y_{\nu}}\bar{\psi}_{in}(z)\frac{\partial\psi_{{\rm in}}(z)}{\partial z^{\rho}}\frac{\partial\psi_{{\rm in}}(z)}{\partial z_{\rho}}\nonumber \\
 &  & +\bar{\psi}_{in}^{2}(x)\bar{\psi}_{in}(y)\frac{\partial\bar{D}^{F}(y-x)}{\partial x^{\mu}}\frac{\partial D^{F}(x-z)}{\partial x_{\mu}}\bar{\psi}_{in}(z)\frac{\partial\psi_{{\rm in}}(y)}{\partial y^{\nu}}\frac{\partial\psi_{{\rm in}}(y)}{\partial y_{\nu}}\frac{\partial\psi_{{\rm in}}(z)}{\partial z^{\rho}}\frac{\partial\psi_{{\rm in}}(z)}{\partial z_{\rho}}\nonumber \\
 &  & +\bar{\psi}_{in}^{2}(x)\bar{\psi}_{in}^{2}(y)\frac{\partial\psi_{{\rm in}}(x)}{\partial x^{\mu}}\frac{\partial D^{F}(x-z)}{\partial x_{\mu}}\frac{\partial\bar{D}^{F}(z-y)}{\partial y^{\nu}}\frac{\partial\psi_{{\rm in}}(y)}{\partial y_{\nu}}\frac{\partial\psi_{{\rm in}}(z)}{\partial z^{\rho}}\frac{\partial\psi_{{\rm in}}(z)}{\partial z_{\rho}}]:.\nonumber
\end{eqnarray}
 \textbf{Note added:} Since the first version of this paper listed on the arXiv, there have been more discussions on the subject
  \cite{Zheltukhin1,Zheltukhin2,kuzenko1,kuzenko2}.
The last two of these showed explicitly equivalences of all these actions.

\begin{acknowledgments}
We would like to thank Simon Tyler for pointing out a typographical
error. This work is supported in part by the National Science Foundation
of China (10425525, 10875103), National Basic Research Program of
China (2010CB833000), and Zhejiang University Group Funding (2009QNA3015). \end{acknowledgments}

\end{document}